\documentclass[10pt, conference]{IEEEtran} 
\IEEEoverridecommandlockouts  

\usepackage[oldenum,olditem]{paralist}
\usepackage{cite}
\usepackage{xcolor}
\usepackage{tikz}
\AtBeginEnvironment{quote}{\quotefont\small}

\usepackage{color, colortbl}
\usepackage{url}
\usepackage{pifont}
\usepackage{placeins}
\usepackage{graphicx}
\usepackage{comment}
\usepackage{mathtools}
\usepackage{float}
\usepackage{amsfonts}

\definecolor{picolor}{RGB}{51,102,0}
\newcommand{\mycheck}[1]{{\color{picolor}{\ding{51}}}}

\newcommand{\ie}{\textit{i.e., }}
\newcommand{\eg}{\textit{e.g., }}

\author{
\IEEEauthorblockN{Nilesh Vyas, Paulo Mendes}
\IEEEauthorblockA{Airbus, Willy-Messerschmitt-Strasse 1, 82024 Taufkirchen, Germany \\
 nilesh.vyas@airbus.com, paulo.mendes@airbus.com }
}

\begin{document}

%\title{Key Management in Quantum Networks}
\title{Relaxing Trust Assumptions on Quantum Key Distribution Networks}

\maketitle
\thispagestyle{plain}
\pagestyle{plain}
\begin{abstract}
Quantum security over long distances with untrusted relays is largely unfounded and is still an open question for active research. Standard Quantum Key Distribution Network (QKDN) architecture demands full trust in QKD relays, which is too restrictive and limits QKDN applications. We investigate secure secret relaying in QKDNs by relaxing the trust assumptions (if not completely) on the relay. We classify QKD relays into Full Access Trust (FAT), Partial Access Trust (PAT), and No Access Trust (NAT) levels, each reflecting the required trust on the key management system for end-to-end communication. We review and propose QKD key management systems based on these trust levels. We propose a new decentralized key management system and review key management within a centralized topology. These different approaches offer various advantages based on QKDN requirements, granting operational flexibility. We believe this work offers a fresh perspective on the challenge of ensuring secure long-range communications in the future.

%As the name suggests, each level defines the degree with which a relay is required to be trusted with the secret provided by the key management system for end-to-end communication. We then review and propose multiple constructions of the QKD key management system based on the different trust levels.
%Main contribution of the paper is realized by evaluating key management systems with no access trust level.
%In principle, we review key management with centralized topology and propose a new decentralized key management system. These different topologies provide various advantages based on the QKDN requirements, allowing an operational flexibility in the architecture. We believe this work presents a new perspective to the open problem of providing a confiding and a practical solution for future long range secure communications.
\end{abstract}

%\begin{IEEEkeywords}
%QKD, Quantum Networks, Key management, Routing.
%\end{IEEEkeywords}

\IEEEpeerreviewmaketitle

\section{Introduction}

\textit{Quantum Key Distribution} (QKD) facilitates the information-theoretic-secure transmission of critical data, securing communication against an eavesdropper with unbounded computational resources \cite{Bennett2014QuantumTossing}. Today, point-to-point QKD systems are already a reality, with commercially available systems performing mostly around one hundred kilometers over optical fibers, and recent academic demonstrations around thousand kilometers \cite{TFQKD}. 

Quantum Key Distribution (QKD) is being studied by various standardization organizations, including the \textit{International Standard Organization} (ISO), E\textit{uropean Telecommunications Standards Institute} (ETSI), and \textit{International Telecommunication Union Telecommunication Standardization Sector} (ITU-T). ISO has established security frameworks and evaluation methods (ISO/IEC 23837-1:2023 and ISO/IEC 23837-2:2023). ETSI focuses on large-scale trusted networks through its Industry Specification Group on QKD (ISG-QKD)\cite{ETSI2023QuantumModules}. ITU-T has issued several recommendations, such as ITU-T Y.3803 for Key Management Systems (KMS) in QKD networks and ITU-T Y.3800, which outlines the conceptual structures of QKD networks, supporting their design, deployment, and operation \cite{ITU-T2020SeriesManagement}.

%Besides commercial systems, QKD is being studied by standardization organizations such as \textit{International Standard Organization}, \textit{European Telecommunications Standards Institute} (ETSI) and the \textit{International Telecommunication Union Telecommunication Standardization Sector} (ITU-T). ISO has released general framework for security requirements, test and evaluation methods for quantum key distribution (ISO/IEC 23837-1:2023 and ISO/IEC 23837-2:2023). ETSI activities aim to target QKD systems for large-scale trusted networks, as is the case of the ETSI Industry Specification Group on Quantum Key Distribution for Users (ISG-QKD) \cite{ETSI2023QuantumModules}. ITU-T has released several QKD recommendations, in particular ITU-T Y.3803 \cite{ITU-T2020SeriesManagement}, for a \textit{Key Management System} (KMS) in a QKD network (QKDN).  ITU-T Y.3800 \cite{ITU-T2020SeriesManagement} provides an overview of the networks supporting QKD. The recommendation aims to provide support for the design, deployment, operation, and maintenance for the implementation of QKDN, in terms of standardized technologies. It also recommends the conceptual structures of a QKDN and a user network.

QKDN's layered structure includes a quantum layer, key management layer, control layer, and management layer. The quantum layer's QKD-modules (transmitter QKD-Tx and receiver QKD-Rx) generate secure keys and communicate over QKD-links with both quantum and classical channels. The key management layer's Key Management System (KMS) handles key distribution. Each QKD-node has QKD-modules and a KMS. The control layer ensures secure and efficient operation, while the management layer oversees Fault, Configuration, Accounting, Performance, and Security (FCAPS) and user network management

%According to the recommendations, the layered structure of QKDN consists of a quantum layer, a key management layer, a QKDN control layer and a QKDN management layer. Within the quantum layer, there exists a QKD-module responsible for generating secure QKD-keys using QKD protocols. This module comprises two fundamental hardware components: a transmitter (QKD-Tx) and a receiver (QKD-Rx), both referred to as QKD-modules. Different QKD-modules communicate with each other over a QKD-link, facilitated by a quantum relay point (such as an optical switch). The QKD-link serves two purposes: a quantum channel to transmit and receive quantum signals securely and a classical channel for data exchange (e.g., time synchronization) between the QKD-modules.

%Key manager or Key Management System (KMS) is a functional module within the key management layer responsible for receiving, managing, relaying, and supplying keys generated by QKD-modules and QKD-links to cryptographic applications. A  QKD-node usually consist a QKD-module(s) and a key manager. The role of the QKDN controller in the QKDN control layer is to oversee QKDN resources, ensuring their secure, stable, efficient, and robust operation. Within the QKDN management layer, the QKDN manager handles Fault, Configuration, Accounting, Performance, and Security (FCAPS) aspects for the entire QKDN, while also supporting user network management.

A key assumption in QKDN \cite{ITU-T2020SeriesManagement, ETSI2023QuantumModules} is that intermediary QKD-nodes or relays must be trusted, implying they are secure against unauthorized access and attacks. Free-space QKD-links to Low Earth Orbit (LEO) satellites have been validated by the Chinese Micius satellite as a trusted courier \cite{Liao2018Satellite-RelayedNetwork}. Trusted nodes function as key relays using "measure and forward" procedures for point-to-point security, decrypting and encrypting secret information with QKD-keys at each relay. As this information is available in plaintext at these relays, they are vulnerable to attacks and need robust security measures.

Providing quantum security over long distance with untrusted relays is still an open question to realise in practice \cite{Huttner2022Long-rangeTechnology}. Extending the QKD range for point-to-point secure communication and overcoming the current fundamental limitations requires sophisticated quantum repeater technology, which is far from practicality. Therefore, quantum networks based on point-to-point QKD-links have been developed using key relay method over trusted nodes. Hence this paper aims to answer the following question: \textit{Is it possible to securely relay secret information in a QKDN by relaxing the trust assumptions (if not completely) on the intermediary relays?} 

We address this question and explore novel constructions of key management systems by relaxing the trust requirements on relays. We believe that the answer to this question presents a new perspective to the open problem of providing a confiding and a practical solution for future long range secure communications.

\section{Related Work}
To explore suitable QKDN solutions, various experiments have been conducted. BBN Technologies \cite{Elliott2005CurrentNetwork}, for instance, developed the first DARPA QKDN with ten quantum nodes, active optical switches, and trusted nodes, using the BB84 protocol \cite{Bennett2014QuantumTossing} to achieve a key rate of 400 bps over 29 km. In 2004, the European Commission's SECOQC project launched a QKDN \cite{Poppe2008OutlineVienna} with six quantum nodes and five different QKD protocols, achieving a key generation rate of 3.1 kbps over 33 km. Both DARPA and SECOQC relied on trusted nodes.

To address untrusted nodes, research has focused on two main approaches: using Measurement Device Independent QKD (MDI-QKD) \textit{Measurement Device Independent QKD} or Twin-Field QKD (TF-QKD) \cite{Lucamarini2018OvercomingRepeaters} to construct hybrid networks \cite{Fan-Yuan2022RobustNodes} of trusted and untrusted nodes (measurement device). However, these solutions still require trusted relays  to extend the QKD distance or multiple disjoint paths to prevent eavesdropping \cite{Cao2019SDQaaS:Service, Zhou2019QuantumManagement}.

Another approach involves quantum satellites, where a satellite generates QKD keys via ground-to-satellite channels with ground-based QKD nodes \cite{Huttner2022Long-rangeTechnology}. The satellite combines the keys using an XOR operation and transmits them over a classical channel. However, this method requires trusting the satellite and can experience deployment delays if using LEO satellites due to orbit time.

\begin{figure*}[hbt!]
    \centering
    \includegraphics[width=\textwidth]{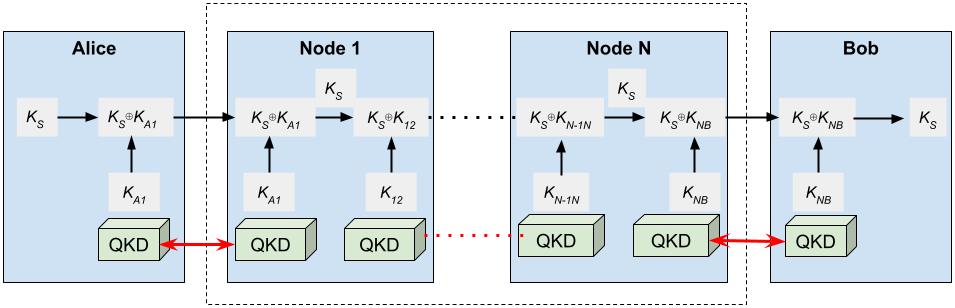}
    \vspace{-1em}
   \caption{QKDN based on FAT relays. Each chained relay decrypt and re-encrypt incoming messages before forwarding it to the next hop.}
    \label{fig:keyforwarding}
\end{figure*}

\section{Trust assumptions in QKDNs}
This paper delves into pioneering methods for securely relaying secret material in a QKDN between distant nodes, Alice and Bob, by reducing the trust requirements on intermediary QKD-nodes or relays. Ideally, deploying large-scale QKDNs with interconnected untrusted relays would be revolutionary for QKD. However, practical demonstrations of such systems remain sparse. Therefore, this paper analyzes various steps to support the progressive relaxation of trust assumptions in current QKD networks.
%This paper explores new techniques to securely relay a secret material in a QKDN between two distant nodes, a sender Alice and a receiver Bob, by relaxing the level of trust required on intermediary QKD-nodes or relays. In an ideal scenario, we should be able to deploy large-scale QKDNs by interconnecting a set of untrusted relays. The development of such a system would represent a game-changer for QKD. However, practical demonstrations of such systems are largely unfounded. Therefore, in this paper we provide an analysis of different steps that may sustain the idea of progressively relaxing the trust assumption of current QKD networks.

\subsection{Honest-But-Curious Adversary}
In our QKDN scenario, Alice and Bob, two distant nodes, wish to share a secret, either a secret-key for encryption or plain text data, via a series of relays. We classify these relays as "\textit{semi-honest}" or "\textit{honest-but-curious adversaries}" \cite{HBC}. They follow the protocol without deviation but seek to learn as much as possible from the messages they legitimately receive. Unlike active or dishonest adversaries who disrupt protocols by deviating from them, honest-but-curious adversaries passively gather information without actively cheating. They might keep records of their intermediate computations, potentially sharing them with other dishonest parties later. This distinction is crucial for understanding the security properties discussed in this paper.
%We consider a QKDN setting where two distant nodes Alice and Bob want to share a secret, for application use, via a series of relays. This secret can be a randomly generated secret-key used for encrypting data at the application or user layer, or can be the data itself in plain text. Each relay is defined as \textit{semi-honest} or \textit{honest-but-curious adversary} \cite{HBC}: a legitimate participant in a communication protocol who will not deviate from the defined protocol, but will attempt to learn all possible information from legitimately received messages. As compared to the active or dishonest adversary, who actively disrupts the protocol by sending messages that deviate from the specified protocol, the honest-but-curious adversary is passive in action and gathers information from legitimately received messages without actively cheating. This could in principle involve the adversary to keep a record of all its intermediate computations which it may later share with the other dishonest parties. This distinction is crucial for understanding security properties in this paper.

\subsection{Trust levels}
This paper characterizes the trust levels on intermediary relays with respect to their capability to access information about shared secret material by learning all possible information from legitimately received messages. The levels of trust are defined as follows:
\begin{itemize}
    \item \textit{Full Access Trust} (FAT): relays have complete knowledge of the secret and need to be fully trusted. This imply the secret is routed through the relay and a record of the same can be stored locally, therefore, the relay needs to be trusted to be secure against unauthorized intrusion and attacks. 
    \item \textit{Partial Access Trust} (PAT): relays do not have complete knowledge, but have partial access to secret keys and so need to be partially trusted. This corresponds to the case where only a part of the secret is being routed through the relay or any computational attempt on the received message reveals only partial information about the secret. Therefore, the relay is required to be trusted to secure this partial information against unauthorized access. 
    \item \textit{No Access Trust} (NAT): relays do not learn any information about secret keys. This implies, either the secret is not routed through the relay or intermediate computation does not outputs any meaningful information about the secret. 
\end{itemize}

\subsection{Key management system constructions}
In our scenario, sender Alice and receiver Bob wish to exchange a secret $K_{S}$, but they lack a direct quantum link and must rely on a chain of $N$ trusted relays. Keys shared between adjacent relays are denoted as $K_{ij}$, for $i \neq j$, $i,j \in \{1, \cdots, N\}$. The key $K_{A1}$ is shared between Alice and her adjacent relay, while $K_{NB}$ is the key between the last relay and Bob.

From a key management perspective, different systems can develop QKDNs with varying trust levels. One example is a decentralized system using routing and forwarding schemes, allowing QKD nodes to exchange secret keys over trusted relays, which decrypt and encrypt the forwarded keys. This forms a FAT QKDN (see Section \ref{sec:FAT}). An evolution of this system is a QKDN that splits a secret key into parts transmitted via multiple paths. This constitutes a PAT system, where relays access parts of the key used for end-to-end communications (see Section \ref{sec:PAT}). For instance, the final secret can be consolidated by XORing all parts received over multiple paths. Each path's nodes must be trusted with their part of the secret, and an adversary must not control all paths in a PAT system.

Beyond FAT and PAT systems, the most desirable QKDN configuration involves NAT relays, which have no or partial access to secret keys used for end-to-end communications (see Section \ref{sec:NAT}). This approach enhances security by minimizing trust in intermediary relays.

\section{QKDNs with Full Access Trust}
\label{sec:FAT}

In a basic networking configuration, distant QKD nodes exchange keys through a QKDN with two or more  full access trust relays. These relays decrypt and re-encrypt the secret $K_{S}$ received from the previous node, using the adjacent QKD keys, before forwarding them. %The secret, encrypted using QKD keys generated over quantum links between adjacent nodes, are safeguarded against eavesdropping on classical channels. %Thus, End-hosts share the secret, however, all relays along the path have a local copy of this secret and needs to be trusted to keep it safe and secure.
%QKDNs with full access trust are built by having each relay, along a lengthy chain, decrypting and re-encrypting received keys before forwarding them to the next hop. Since forwarded keys are encrypted using QKD keys generated over the quantum links connecting adjacent nodes, they are protected from eavesdropping over classical channels. In this process, end-hosts share the QKD key generated by one of them over its local quantum link, or by using a local quantum device. However all relays are acquainted with the transmitted key, which is used to secure the communication between end-hosts.
As illustrated in figure \ref{fig:keyforwarding}, key forwarding between two nodes, Alice and Bob, that are not directly connected via a quantum link, requires, in a basic approach, the following set of operations:
 \begin{enumerate}
         \item Alice generates a key, $K_{S}$, to be shared with Bob via relays;
         \item Alice uses $K_{A1}$ to encrypt the key $K_{S}$ as \begin{math} K_{S} \oplus K_{A1} \end{math}, and send it to the Node 1;
         \item  Node 1 uses $K_{A1}$ to decrypt the received message and acquires key $K_{S}$;
         \item  Node 1 encrypts $K_{S}$ using $K_{12}$ shared with Node 2 as \begin{math} K_{S} \oplus K_{12} \end{math}, and send it to Node 2;
         \item Next relays perform the same operations as Node 1, until Bob is reached.
         \item Bob decrypts the received message,  \begin{math} K_{S} \oplus K_{NB} \end{math}, using $K_{NB}$ and acquires the key $K_{S}$.
\end{enumerate}

In this QKDN setup, relays can fully access the secret $K_{S}$, used by end-hosts for encryption. This exemplifies a QKDN with a Full Access Trust (FAT) level. Here, the security of the secret during transmission is ensured by XOR operations, protecting it against chosen plain-text attacks. However, since all relays are acquainted with the transmitted secret, they must all be trusted. From a network management perspective, this is problematic because it’s hard to quantify the security risk and relate it to business assets (confidential information) exchanged over the quantum network. Despite this, the FAT transmission model is gaining traction and is being implemented in various quantum network deployments.

%In this QKDN setting, relays are able to learn the complete information about the secret $K_{S}$, that will be used by end-hosts to encrypt their communication. Therefore this system is an example of a QKDN with a FAT level. In this trust level, the security of the secret during its transmission is guaranteed by XOR operations, preserving security of the secret against chosen plain-text attacks (IND-CPA).

%This solution however allows all relays to be acquainted with the transmitted secret, meaning that all relays need to be trusted. From a network management perspective, such assumption is rather undesirable, since it is difficult to quantify the security risk and hence meaningfully relate it to business assets (confidential information) that are exchanged over the quantum network. Nevertheless, this transmission paradigm is gaining attention and finds itself implemented in various deployments of quantum networks.

\begin{figure*}[hbt]
    \centering
    \includegraphics[width=\textwidth]{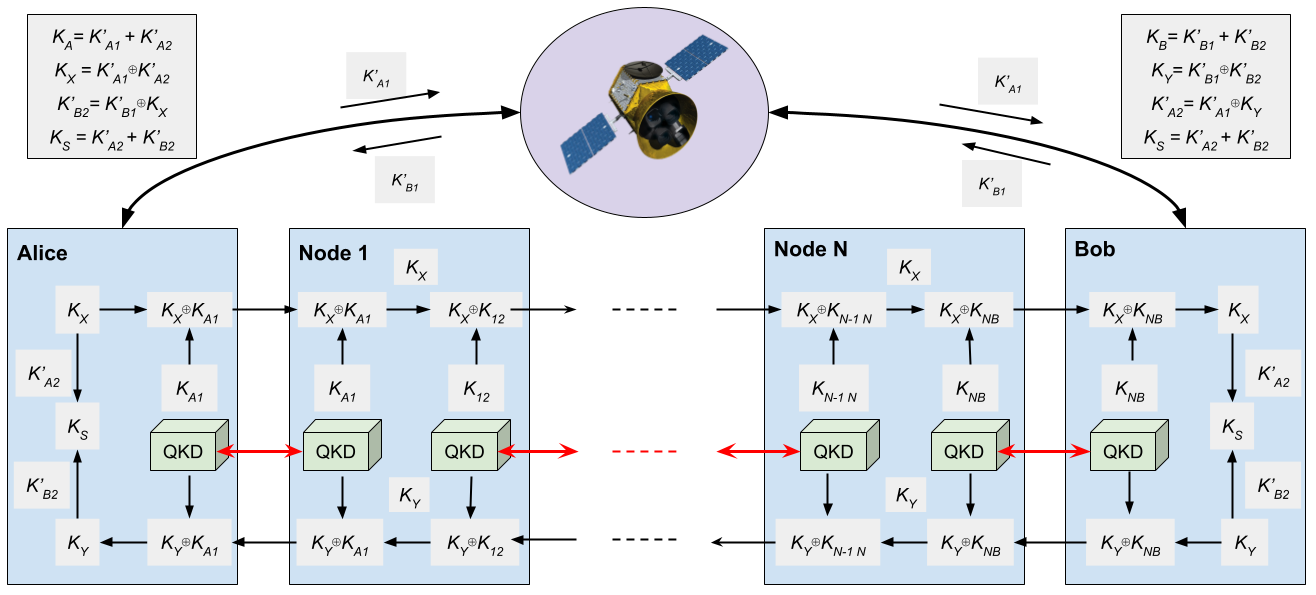}
    \vspace{-1em}
    \caption{Decentralized KMS for QKDN with partial-access trust on the quantum nodes}
    \label{fig:satellite-enabled-QKD}
\end{figure*}

\section{QKDNs with Partial Access Trust}
\label{sec:PAT}

Using multiple paths for transmission is an effective way to address the stringent requirements of QKDNs with FAT trust levels. This method was initially proposed in \cite{multipath1} and further explored in \cite{Multipath2, Multipath3}. In the simplest setup, Alice and Bob, two QKD nodes, share a secret by splitting it into multiple parts, with each part sent through different paths (as detailed in section \ref{sec:FAT}). We consider two scenarios of how the secret can be split and reconfigured at the Bob's end:

%Transmission over multiple paths is a straightforward remedy to relieve the stringent requirement of QKDNs with FAT trust level. This approach has been proposed in \cite{multipath1} and recently explored in \cite{Multipath2, Multipath3}. In the simplest setting, two QKD nodes, Alice and Bob, share a common secret over multiple paths. This is achieved by having Alice splitting the secret into multiple parts, sending each one of them though different paths in a similar fashion as described in section \ref{sec:FAT}. 

\begin{itemize}
    \item \textit{Using Secret Sharing}: The decomposition of the secret key into multiple parts is an example of secret sharing. A simple skeleton of multi-path transmission can be derived from the well-known polynomial $(t, k)$ secret-sharing due to Shamir (1979) \cite{Shamir}. The aim of such scheme is to divide a secret $K_{S}$ into $k$ shares or parts, $K_{Si}$ for $i \in \{1,\ldots , k\}$, such that, (a) knowledge of any $t$ or more shares $K_{Si}$ are required to make $K_{S}$ easily computable; (b) knowledge of any $t-1$ or fewer $K_{Si}$ parts leaves $K_{S}$ completely undetermined or all its possible values are equally likely. 
The secret can be shared in the following way,
%To share a secret $K_{S}$, %Alice now chooses a random polynomial $f(x)$ of degree $t-1$ such that $f(0)=s$, and sends $f(i)$ over each of the  $i_{th}$ path for $i \in \{ 1,\cdots, k\}$. To reconstruct the secret from sufficiently many shares, the polynomial can be reconstructed, and evaluated at 0. More formally,
\begin{enumerate}
    \item Alice chooses the secret $K_{S}$, where $K_{S}\in \mathcal{S}:= \mathbb{Z}_{q}$. 
    \item Alice split $K_{S}$ as follow; choose $t-1$ numbers randomly from $\mathbb{Z}_{q}$ \ie $a_{1}, \ldots a_{t-1} \xleftarrow{\$ }\mathbb{Z}_{q}$ such that $a_{t-1}\neq 0 $, and define
    \begin{equation}
        f(x) := a_{k-1}x^{k-1} + \cdots +  a_{1}x + K_{S}
    \end{equation}
    the algorithm now outputs each share as $K_{Si}=(i, f(i))$ for $i \in \{ 1,\cdots, k\}$.
    \item Alice sends each $K_{Si}$ over $k$ paths.
    \item Relays on each path follows the protocol as described in section \ref{sec:FAT}.
    \item Bob on receiving all $k$ shares, chooses any $t$ shares to compute a unique interpolation polynomial $g(x)$ of degree $t-1$ in $\mathbb{Z}_{q}[x]$ \cite{Shamir} and reconstruct the free coefficient $g(0)$ as the $K_{S}$. 
\end{enumerate}

Given $k$ disjoint paths, an adversary’s optimal strategy is to attack only one relay node per path \cite{Multipath3}. Consequently, the security of this scheme necessitates that no more than $t-1$ disjoint nodes in the network can be compromised. Thus, $t$ represents the adversary’s threshold.%This scheme has another advantage from recovery against packet loss in the network, similar to what network coding is used for in the classical networks. As only, $t$ shares are required to reconstruct the secret, this scheme can withstand upto $k-t$ packet loss.  
    \label{secretsharing}
    
    \item \textit{Using XOR Operations}: The final secret can be consolidated by XORing all the parts received over $k$-paths. In such a setting, Alice chooses $k$ strings $K_{Si}$ for for $i \in \{ 1,\cdots, k\}$, of equal lengths and compose the secret as 
\begin{equation}
    K_{S} = K_{S1}\oplus\cdots\oplus K_{Sk}
    \label{secretmulti}
\end{equation}
Security and performance analysis of such scheme has been analysed for disjoint and overlapping paths in \cite{Multipath3}. For disjoint paths, each relay in every path has partial knowledge $K_{Si}$ over the secret and therefore it must be trusted against the unauthorized access to the part of the secret, otherwise this could be considered as a vulnerability of the system as information leakage. Moreover, an adversary does not gain any information on the secret as long as at least one of the path is secured against unauthorized access.
    \label{XORoperation}
\end{itemize}
%\subsection{Using XOR Operations}
As long as an adversary cannot control all paths, the security of the secret can be ensured. The adversary can only learn partial information about the secret by controlling up to $t-1$ paths in the first example (\ref{secretsharing}) and $k-1$ paths in the second example (\ref{XORoperation}). Thus, multipath key forwarding satisfies the PAT level, as relays can access only partial information of the secret flowing though its path.

\subsection{Decentralized Key Management System}
\label{sec:decentralized}
We propose a decentralized Key Management System (KMS) to support QKDNs utilizing a set of PAT relays. As depicted in Figure \ref{fig:satellite-enabled-QKD}, this decentralized QKDN implementation involves dual-path and two-way communication between Alice and Bob: one path via terrestrial relays and another via a satellite. Each intermediate relay node and the satellite host a local KMS, facilitating two-way communication.

The secret key is formed by XORing random bit strings generated locally by Alice and Bob. Instead of transmitting these random bit strings directly over the network, an XOR-encrypted version of the bit string is sent via the terrestrial network, while the XOR encryption key is transmitted via the satellite. This dual-path scheme's security hinges on the assumption that the two paths remain distinct and do not intersect.

%Here we present a decentralized KMS aim to support QKDNs based on a set of PAT relays. Figure \ref{fig:satellite-enabled-QKD} illustrates a possible implementation of a decentralized QKDN. The architecture leverages a two way communication between Alice and Bob over two paths; first over a set of terrestrial relays and second using a satellite. Key management system is decentralized by having a local KMS at each intermediate relay node and the satellite, facilitating a two way communication. 

%The secret key is composed by XORing the random bit strings generated locally at the Alice and Bob side. Moreover the random bit strings are not forwarded directly over the network, instead an XOR encryption of the random bit string is forwarded over the terrestrial network and the XOR encryption key is routed over the satellite. Thus, the security of this scheme relies on the assumption that the two paths do not intersect.

To avoid intermediate relays for gathering information about the shared key \( K_{S} \), a hybrid approach can be employed as follows:
\begin{enumerate}
    \item Alice and Bob each choose a random bit string, \( K_{A} \) and \( K_{B} \), respectively, and split them into two equal parts:
    \begin{equation}
        K_{A} = K_{A1}^{'} + K_{A2}^{'}, \quad K_{B} = K_{B1}^{'} + K_{B2}^{'}
    \end{equation}
     \item They generate bit strings \( K_{X} \) and \( K_{Y} \) by performing the XOR operation on the split strings:
    \begin{equation}
       K_{X} = K_{A1}^{'} \oplus K_{A2}^{'}, \hspace{0.5cm} K_{Y} = K_{B1}^{'} \oplus K_{B2}^{'}
    \end{equation}
    \item Alice sends \( K_{X} \) to Bob and Bob sends \( K_{Y} \) to Alice over a series of \( N \) intermediate QKD relays, as described in Section \ref{sec:FAT}.
    \item They forward \( K_{A1}^{'} \) and \( K_{B1}^{'} \) to each other via satellite and store \( K_{A2}^{'} \) and \( K_{B2}^{'} \) locally.
    \item Upon receiving \( K_{Y} \) and \( K_{B1}^{'} \), Alice performs XOR to obtain \( K_{B2}^{'} \). Similarly, Bob retrieves \( K_{A2}^{'} \) from \( K_{X} \) and \( K_{A1}^{'} \).
    \item Finally, the the secret key is composed as
    \begin{equation}
        K_{S} = K_{A2}^{'} +  K_{B2}^{'}.
    \end{equation}
\end{enumerate}

Alice and Bob generate random strings \(K_{A}\) and \(K_{B}\) locally. The segments 
\(K_{A2}^{'}\) and \(K_{B2}^{'}\), which form the secret key  \(K_{S}\), are never transmitted directly. Instead, XOR encryptions  \(K_{X}\) and \(K_{Y}\) are sent via relay nodes, while \(K_{A1}^{'}\) and \(K_{B1}^{'}\) are sent via satellite. If the paths are disjoint, intermediate QKD relays can't access \(K_{S}\), but they must be trusted with the routed information, reducing the trust requirement to the PAT level. 

A cost-effective QKDN can be built with terrestrial QKD relays and a Geostationary Orbit (GEO) satellite. Keys can be transported via secure protocols like QUIC or TCP/TLS over the satellite link \cite{Dierks2008TheProtocol}. In such decentralized QKDN, Alice can initiate end-to-end communication with Bob via a local application. They should establish a transport connection (TCP/TLS or QUIC) over the satellite link and create a local key pool for application use. The key pool size should be based on metrics negotiated during the transport connection setup. For long-term security, especially against quantum adversaries, GEO satellites can be upgraded to LEO or MEO constellations with post-quantum cryptography (PQC) capabilities or QKD payloads, ensuring cryptographic agility.

\begin{figure*}[hbt!]
    \centering
    \includegraphics[width=\textwidth]{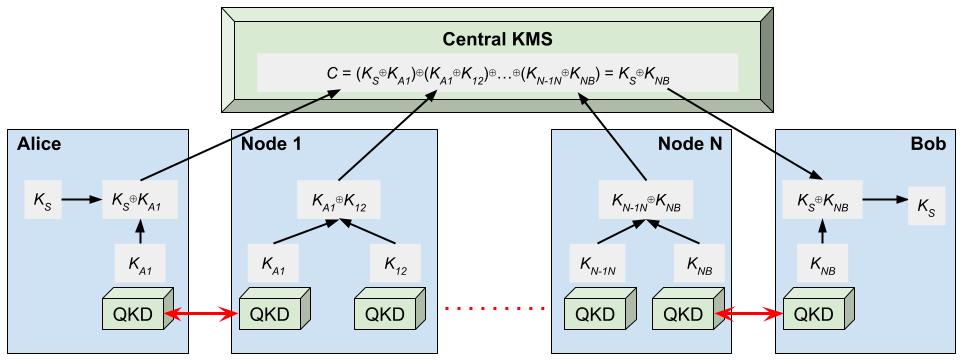}
    \vspace{-1em}
    \caption{Centralized KMS for QKDNs with NAT relays not acquainted with secret keys $K_{S}$ shared between Alice and Bob via the central KMS.}
    \label{fig:CentralKMSr}
\end{figure*}

\section{QKDNs with no Requirements for Access Trust}
\label{sec:NAT}

This section aims to discuss methods to develop QKDNs with no requirements for access trust. In this regard we disucss a Centralized KMS\footnote{https://securecommunications.airbus.com/en/news/quantum-key-distribution-qkd-networks-key-management}. 

\subsection{Centralized Key Management System}
\label{sec:centralized}
A Centralized Key Management System (KMS) features several chained relays connected to a central key manager, a logically centralized node. Each relay connects to two adjacent neighbors, sharing a pair of QKD keys. Relays use these QKD keys to create local secure encrypted masks. Specifically, each relay performs a XOR operation on the two QKD keys shared with its neighbors, while Alice XORs the secret  $K_{S}$ with the QKD key shared with her adjacent node. The encrypted masks generated by the relays (except Bob) are sent to the central key manager over a classical channel, assumed to be secure from eavesdropping during creation and submission. The central KMS then XORs all received masks and shares the output with Bob, enabling him to decrypt and receive the secret  $K_{S}$, as illustrated in figure \ref{fig:CentralKMSr}. 

The \textit{modus operandi} for exchanging the secret between Alice and Bob requires the central key manager to compute on all the keys shared by each relay. The centralised secret key forwarding system encompasses the following steps: 
\begin{enumerate}
    \item Alice chooses a secret $K_{S}$ and generate a mask by XORing it with the shared QKD key adjacent Node 1: $K_{S} \oplus K_{A1}$ 
    \item Each intermediate QKD relay node between Alice and Bob,  creates a mask by computing XOR operation on adjacent shared QKD keys. Node 1: $K_{A1} \oplus K_{12}$, Node i : $K_{(i-1)i}\oplus K_{i(i+1)}$ for $i \in \{2,\ldots, N-1\}$; Node N: $K_{N-1N} \oplus K_{NB}$.
    \item Alice and each relay sends the computed mask to the central key manager. 
    \item Central key manager computes XOR on all masks shared by the relays \ie,
    \begin{eqnarray}
        C &=& (K_{S}\oplus K_{A1}) \oplus (K_{A1}\oplus K_{12}) \oplus \cdots \\
         & & \oplus (K_{(N-1)N}\oplus K_{NB})  \nonumber \\
        &=& K_{S} \oplus K_{NB} 
    \end{eqnarray}
    \item Central key manager sends the output $C$ to Bob.
    \item Bob performs a XOR using $C$ and the shared QKD key $K_{NB}$ to obtain the $K_{S}$.
\end{enumerate}
The relays are not provided with secret key $K_{S}$ that is shared between Alice and Bob for end-to-end communication, which is only relayed through the central KMS. During the transmission from Alice to the central KMS, the security of the secret $K_{S}$ is guaranteed by the XOR operations with the QKD key $K_{A1}$. The XOR computation performed by the central KMS on all shared keys results in $K_{S} \oplus K_{NB}$, which does not revel any information about the secret $K_{S}$, as $K_{NB}$ is unknown to the central KMS. Finally, XOR operation with $K_{NB}$ also protects the transmission of $K_{S}$ from central KMS to Bob. Since no secret information for end-to-end encryption is exchanged through the relays, trust requirements for relay nodes are relaxed to the No Access Trust (NAT) level.

However, this centralized architecture has a single point of failure—the central KMS. If the central KMS fails, the entire QKDN becomes non-operational. Therefore, the central key manager must remain highly secure to ensure the network's reliability. Moreover, the \textit{Centralized  KMS} relies on a shared communication medium (\eg satellite link) to send secret material from several relays to a central device, a satellite, via a wireless medium that may need to be protected. If a relay is able to eavesdrop the communication over the shared wireless medium, it can in principle obtain the secret key and the KMS setting will no longer offer NAT trust level. A computationally unbounded adversary with capabilities to spoof the network traffic can easily impersonate the central KMS and obtain all messages. 

While the assumption that classical communication between the central KMS and nodes is secure from eavesdropping is strong, it is necessary for the scheme's security. The practical limitations of a centralized approach can be mitigated by securing the communication channels between the central key manager and each QKD node (relays and end-hosts). This can be achieved by using a QKD-enabled satellite to protect transmissions between the satellite and QKD nodes with a QKD key. Alternatively, Post-Quantum Cryptography (PQC) can be used to encrypt the channels between the QKD nodes and the central KMS. Both methods protect secret material against quantum adversaries.

Using a QKD-enabled satellite provides an information-theoretic security (ITS) solution, maintaining the ITS quality of the QKDN. However, this approach is expensive and not scalable currently due to the significant hardware burden it places on the central key manager. Conversely, the PQC solution is cost-effective and scalable with simple implementation, but it only offers computational security and requires large storage, thereby challenging the ITS security of the QKD

\section{Conclusion}
In this work, we have investigated the possibility to securely relay secret information in a QKDN by relaxing the trust assumptions (if not completely) on QKD relays. To address this issue, we characterized the trust assumptions on relays into three different levels, namely \textit{Full Access Trust} (FAT), \textit{Partial Access Trust} (PAT), and \textit{No Access Trust} (NAT). As the name suggests, each level is defined by the ability of an QKD relay to access information provided at the key management layer of the QKDN stack, i.e., the secret key used for end-to-end communication. 

We explore various constructions of a key management system considering different trust levels. 
%Striking feature of the work is the QKDN construction with no access trust requirement on QKD relays. 
We describe two different QKD key management frameworks, based on centralized and the decentralized topology, respectively. These different topology provide various advantages based on the QKDN requirements, allowing an operational flexibility in the architecture.

%However, in order to implement all the mentioned security and indirection methods, a decentralized QKDN may induce a high delay in provisioning the security keys requested by applications. So, there is the need to evaluate how the system can provide QKD keys fulfilling the key rate requested by applications, while ensuring a good usage of resources needed for the local key pools. 

In reality, a practical QKDN will be complex system with arbitrary topology, which may benefit from a hybrid solution for relaying the secret keys securely between two distant end hosts. In principle, a hybrid key management system can be designed by considering a partially centralized and partially decentralized topology. For instance, a central KMS supporting key management in metropolitan areas under the control of specific entities (\eg local governments), being such areas connected via a decentralized KMS allowing the secure routing of the secret keys over long distances. 

We believe that this work presents a new perspective to the open problem of providing a confiding and a practical solution for future long range secure communications based on QKD technology.

\ifCLASSOPTIONcaptionsoff
  \newpage
\fi

\bibliographystyle{IEEEtran}
\bibliography{bibliography.bib}
\end{document}